\documentstyle[12pt,epsf]{ioplppt}

\def\asigma{\stackrel{\leftrightarrow}{\sigma}}
\def\no{\noindent}
\def\gs{\gtrsim}
\def\ls{\lesssim}
\def\be{\begin{equation}}
\def\en{\end{equation}}                  
\newcommand{\bi}[1]{\mbox{\boldmath$#1$}}
\newcommand{\av}[1]{\langle{#1}\rangle}

\def\gdot{\dot{\gamma}}
\def\gs{> \kern -12pt \lower 5pt \hbox{$\displaystyle{\sim}$}}
\def\ls{< \kern -12pt \lower 5pt \hbox{$\displaystyle{\sim}$}}
\def\bea{\begin{eqnarray}}
\def\ena{\end{eqnarray}}
\def\p{\partial}

\begin{document}
\title{Nonlinear Strain Theory of Plastic Flow in Solids}
\author{Akira Onuki}
\address{Department of Physics, Kyoto University, 
Kyoto 606-8502, Japan}


\begin{abstract}
We present a phenomenological 
 time-dependent Ginzburg-Landau theory of 
 nonlinear plastic deformations 
in solids.  Because the problem is very complex,
we first give models 
in  one and two dimensions 
without vacancies and interstitials, where 
large strains produce densely distributed slips 
but the mass density deviations remain small except near 
the tips of slips. Next 
we set up a two-dimensional model 
including  a  vacancy 
field (or local free-volume fraction), 
where  relevant is the sensitive  dependence of 
the elastic shear modulus 
on the vacancy density. In our simulation, if   
strains are applied  to   nearly 
defectless solids but in the presence of such 
elastic inhomogeneity,  
the vacancy density and the mass density 
can become considerably heterogeneous for large strains 
on spatial scales much longer than the atomic size. 
These strain-induced disordered states 
are  metastable or long-lived 
once they are created.

\end{abstract}

\section{Introduction}

Most previous papers on glass transitions 
so far are concerned with near-equilibrium 
properties such as  relaxations of the density time correlation 
functions or dielectric response.   
However, these quantities 
are too restricted or indirect, and 
there remains a rich group of unexplored problems 
in far-from-equilibrium  states.  
In particular,  shear 
   is a relevant perturbation 
drastically changing the glassy  dynamics when  
the shear rate $\gdot$ exceeds the inverse of 
the structural relaxation time 
$\tau_{\alpha}$  \cite{Yamamoto_Onuki,Onukibook}. 
In  supercooled liquids,   
the microscopic  rearrangement processes 
occur on the time scale of $\tau_{\alpha}$ in quiescent states, 
whereas  they are much accelerated 
even by extremely small $\gdot$ (if larger than 
$\tau_\alpha^{-1}$).  
Similar {\it jamming rheology}  
has been studied   in   systems
composed of  large 
elements such as colloidal suspensions, 
dense microemulsions, and granular materials \cite{Liu}.

Experimentally, 
Simmons {\it et al}.  \cite{Simons} observed  shear-thinning behavior, 
where the steady-state viscosity $\eta (\gdot)$ 
was  represented by   
\be
\eta (\gdot)
=\sigma_{xy}/\gdot \cong \eta (0)/(1+\gdot \tau_{\eta} ), 
\label{eq:1}
\en
 in the range 
$7\times 10^5 < \eta (0) < 6 \times 10^{13}$ (Poise) 
  in  soda-lime-silica glass.  
The characteristic time $\tau_{\eta}$ is expected to be  
 of order $\tau_\alpha$.  Remarkably,  the shear stress 
$\sigma_{xy}$ tends to a  limiting shear stress,  
$\sigma_{\rm lim}= \eta (0)/ \tau_{\eta}$, of order $10^{-2}\mu_0$, 
$\mu_0$ being the shear modulus for infinitesimal strains.  
After application of shear,  they also observed 
an overshoot of the shear  
stress before approach to a steady state.

As a closely  related problem,   
understanding  of mechanical properties 
of amorphous metals has  been  
of great technological importance \cite{HSChen,theory}. 
They are usually ductile 
in spite of their high strength.  
At low temperatures $T \ls~ 0.6 \sim 0.7 T_g$,  
localized shear bands, 
where  zonal slip occurs,  have been observed 
above a yield stress. At relatively high temperatures 
$T \gs~ 0.6 \sim 0.7 T_g$, on the other hand, 
shear deformations are induced 
{\it homogeneously} (on macroscopic scales) throughout samples, 
giving rise to  viscous flow with strong shear 
thinning behavior. In a model amorphous metal in 3D,  
Takeuchi {\it et al.} \cite{Takeuchi} 
numerically 
followed atomic motions after application of a small shear strain to observe 
 heterogeneities among  poorly and closely  
 packed regions. Such dynamic heterogeneities 
 have  been reported in recent simulations 
 in  sheared states   \cite{Yamamoto_Onuki}
 and also in quiescent states  
 \cite{Muranaka,Harrowell,Yamamoto_Onuki1,Donati}.

In the study of  elasto-plastic 
dynamics of solids, microscopic simulations  
 are  informative \cite{Argon,Langer,Ikeda}, while 
mean-field  theories  
are  instructive  \cite{Doi,Sollich}.     
In the latter theories, the problem was  reduced  to 
to that of one element obeying a 
stochastic process under the influence of the average stress. 
The  aim of  this paper is then 
to present a  space-time dependent,  elasto-plastic  
  dynamical model  on the basis of well-defined 
nonlinear elastic theory.   
Although our theory is still preliminary, 
we shall see that introduction of 
an order parameter, representing the 
vacancy or the local free volume,  can 
give rise to dramatic effects 
 similar to those reported 
in Ref.\cite{Argon}.

\section{Plastic flow in one dimension}

To introduce the fundamental concepts in our 
problem in the simplest manner,  
we first present a  one-dimensional model which mimics 
a solid with  shear  deformations 
varying only in one direction (1D slip model). 
We here write down 
the dynamic equations in the continuum 
representation (for simplicity).  
The velocity  $v(x,t)= \p  u(x,t)/\p t$  
of the  shear displacement  $u(x,t)$  (along the $y$ axis) is 
 governed by 
\be 
\rho \frac{\p}{\p t}v =  \nabla_x [\mu_0 \gamma_0 \sin( 
{\gamma_0}^{-1}{\nabla_x}u)  ] 
 +\eta_0 \nabla_x^2v + \nabla_x \sigma_{\rm R}  ,    
\label{eq:2}
\en 
where $\nabla_x=\p/\p x$, $\rho$ is the mass density, 
 $\mu_0$ is the shear modulus, $\gamma_0$ is a constant 
 determined by the underlying crystal structure, 
 and $\eta_0$ is a vicosity. The period of the strain 
 is given by $\gamma_{\rm p}= 2\pi\gamma_0$. To ensure 
 equilibrium in the absence of  applied force, we introduce the 
  random stress   $\sigma_{\rm R}(x,t)$  
related to $\eta_0$ via 
the fluctuation-disipation relation 
$\av{\sigma_{\rm R}(x,t)\sigma_{\rm R}(x',t')} 
= 2k_{\rm B}T \eta_0 \delta (x-x') \delta (t-t)$, where $T$ is the temperature. 
Hereafter   space and time will be measured  in units of $\ell\equiv  
 \eta_0/(\rho \mu_0)^{1/2}$ 
 and $\omega_0^{-1} \equiv  \eta_0/\mu_0$, respectively. 
 The  dimensionless 
 noise strength is given by  
   $\epsilon= k_{\rm B}T/\mu_0\gamma_0^2 \ell$.

We numerically solve (2) on a 1D chain  
($j=0,1,\cdots,N=600)$ with $\epsilon=0.25$,    setting  
$u(x,t) \rightarrow u_j(t)$ 
and  $\p u/\p x 
\rightarrow \gamma_j\equiv u_{j+1}- u_j$. 
  We apply a constant 
shear rate $\gdot$ at $t=0$  with 
$u_0=0$ and $u_N/\gamma_0= N\gdot t$. 
In Fig.1 the average dimensionless stress $\av{\sigma}(t)=
N^{-1}\sum_j \sin 
(\gamma_j(t)/\gamma_0)+ \gdot$ is plotted as a function of 
 the scaled strain 
$\gdot t/\gamma_{\rm p}$.   The 
system undergoes a pure elastic deformation 
$\gamma_j(t) /\gamma_0 \cong \gdot t$ in the initial stage,
   but   slips (jumps of 
$\gamma_j$ by multiples of $\gamma_{\rm p}$) appear 
randomly throughout the system with increasing 
the average strain $\gdot t$.  The  inset 
shows   slips and 
continuously strained  regions. The average  slope of 
the latter regions gives the  average  stress. 
For the range of the shear rates in Fig.1, we 
 can see shear-thinning   
$\sigma_{xy} \propto \gdot^{0.7}$ in steady states 
(if averaged over time).

It is also straighforward to integrate (2) under a fixed 
shear stress $\sigma_{\rm ext}$  applied 
at one end with the other end being pinned. 
For small $\sigma_{\rm ext}$  below a yield stress 
$\sigma_{\rm y}$, we observe no slip formation, 
while for  $\sigma_{\rm ext}> \sigma_{\rm y}$ 
plastic flow is produced.  For the noise strength 
of $\epsilon=0.25$, we find 
$\sigma_{\rm y} \cong 0.2$.

\section{Plastic flow without vacancies}

 As a direct generalization of (2), 
we  set up a plastic flow  model in two dimensions  
(2D slip odel).  
In the continuum limit  the 
strain components are defined by 
\be 
e_1=\nabla_x u_x+ 
\nabla_yu_y, \quad e_2= \nabla_x u_x- 
\nabla_yu_y, \quad e_3= \nabla_x u_y 
+\nabla_yu_x, 
\label{eq:3}
\en 
where $\nabla_x=\p/\p x$ and $\nabla_y=\p/\p y$.
If we suppose a triangular lattice,  
the elasticity is isotropic in the harmonic 
approximation \cite{Onukibook}, being characterized  
by the bulk and shear moduli,  $K_0$ and $\mu_0$, 
but it depends on the orientational angle $\theta$ 
of one of the crystal axes for large shear deformations. 
Note that, under rotation of the reference frame 
by $\theta$, the shear strains $e_2$ and $e_3$ are changed 
to  $e_2'$ and $e_3'$, where \cite{Onukibook}
\be 
e_2'= e_2 \cos 2\theta + e_3 \sin 2\theta, \quad 
e_3'= -e_2 \sin 2\theta + e_3 \cos 2\theta.  
\label{eq:4}
\en
We write    the  elastic energy density  
in the form, 
$f_{\rm el}= {K_0}e_1^2/2 + \mu_0 {\cal F}(e_3', e_2')$.
The simplest form of  the scaling function ${\cal F}$ 
is of the form   
\be 
{\cal F}(e_3', e_2')= \frac{1}{6\pi^2} 
\bigg [ 3- \cos\pi(\sqrt{3}e_3'-e_2') - \cos\pi(\sqrt{3}e_3'+e_2')
-  \cos(2\pi e_2') \bigg ] 
\label{eq:5}
\en  
and is shown in Fig.2, The 
$\cal F$ is invariant with respect to the rotation 
$\theta \rightarrow  \theta +  \pi/3$,  
is a periodic function of $e_3'$  
with period $\gamma_{\rm p}=2/\sqrt{3}$ for $e_2'=0$ 
(simple shear deformation),  and 
becomes $(e_2^2+e_3^2)/2$ for small strains.  
If we assume $\p \theta/\p t= 
(\nabla_x v_y- \nabla_y v_x)/2$ for the angle rotation rate, 
its time integration yields 
\be 
\theta = \frac{1}{2} (\nabla_x u_y- \nabla_y u_x) +\theta_0 , 
\label{eq:6}
\en 
where $\theta_0$ is  independent of $t$ but 
may depend on ${\bi r}=(x,y)$. 
For $\theta=0$ one of the crystal axes is along the $x$ axis.  
Then the elastic energy $F_{\rm el}= \int d{\bi r}f_{\rm el}$ 
is a functional of $\bi u$. 
We assume that the lattice velocity ${\bi v}= \p {\bi u}/\p t $ obeys 
\be
\rho \frac{\p}{\p t}{\bi v} =  
-\frac{\delta}{\delta {\bi u}} F_{\rm el}  
 +\eta_0 \nabla^2v + \nabla \cdot \asigma_{\rm R}  ,    
\label{eq:7}
\en 
where   
 the first term on the right hand side 
is also written as $\nabla\cdot\asigma$ in term of the 
elastic stress tensor $\asigma$.  
The symmetric random stress tensor $\asigma_{\rm R} =
\{\sigma^{\rm R}_{\alpha\beta}\}=  
\{\sigma^{\rm R}_{\beta\alpha}\}$ 
satisfies $\sigma^{\rm R}_{xx}+\sigma^{\rm R}_{yy}=0$, because the bulk 
viscosity is neglected in (7), and \cite{Onukibook} 
\be 
\av{\sigma^{\rm R}_{\alpha\beta}({\bi r},t)
\sigma^{\rm R}_{\alpha\beta}({\bi r}',t')}=
2k_{\rm B}T \eta_0 \delta({\bi r}-{\bi r}')\delta(t-t').  
\label{eq:8}
\en

We measure  space and time in units of 
$\eta_0/(\rho \mu_0)^{1/2}$ 
 and $\eta_0/\mu_0$   and the strains in units 
 of $\gamma_{\rm p}= 2/\sqrt{3}$. 
If  these  scaling units are used, the noise strength 
$k_{\rm B}T\eta_0$ in (8) is replaced by 
\be 
\epsilon =  k_{\rm B}T \rho/\gamma_{\rm p}^2 \eta_0^2.
\label{eq:9}
\en  
We   integrate (7) on a $128 \times 128$ square lattice 
by applying a constant shear rate $\gdot$ at $t=0$ 
with $\epsilon=0.1$.
The periodic boundary condition is  imposed in the 
$x$ direction, while  
${\bi u}= {\bi 0}$ at the bottom $y=0$ 
and $u_x=\gdot t$ 
and $u_y= 0$ at the top $y=128$.  
At $t=0$, the values of $\bi v$  at the lattice sites 
are Gaussian random numbers 
with variance $\epsilon^{1/2}$ 
 but  ${\bi u}={\bi 0}$ and $\theta=0.1$.
In Fig.3 we show the average scaled stress 
$\av{\sigma_{xy}}/\mu_0$ as a function of  the scaled strain 
$\gdot t/\gamma_{\rm p}$ for $\gdot=10^{-3}$ and $10^{-4}$. 
In the inset we display snapshots of $e_3$ and $e_2$.  
Because  we start with  a perfect crystal with 
the initial fluctuations 
only in the velocity,  the stress curve drops sharply 
after the peak with catastrophic formation of slips. 
Then  structurally disordered 
states  are produced 
where defects are proliferated ({\it strain-induced 
disordering}). Note that slips are accumulations of 
 dislocations $[17]$.  In the initial stage of plastic flow 
we have simple slips consisting of two dislocations with opposite 
Burgers vectors $\pm a$ with $a$ being the lattice constant. 
The elastic energy to create such 
a slip is minimum in the $x$ and $y$ directions under 
shear deformation.  
This is the reason why the slips in Fig.3 are 
parallel to the $x$ or $y$ direction.  
In Fig.4  the  shear $(=10^{-3})$  is  swiched off 
at  (a) $t=190$ before the peak time of the stress, 
(b) $t=220$ just after the peak time, and 
(c) $t=440$.  The top and bottom 
boundaries are kept at rest afterwards. 
Pure elastic deformation is maintained 
in (a), while 
no appreciable time evolution is detected  after 
transients in (b) and (c). 
This means that the structurely 
disordered  states are metastable.

\section{Plastic flow with vacancies}

Usually in the literature, the density deviation 
$\delta\rho$ is equated with $-{\bar \rho} \nabla\cdot{\bi u}= -
{\bar \rho} e_1$ 
in solids \cite{Landau},  where 
 $\bar{\rho} (\cong \rho)$  is the average 
density  and $|\delta\rho|\ll \bar{\rho}$ 
is  assumeed.    In the presence 
of vacancies or interstitials, however, there  can be 
a small difference between these two quantities. 
Cohen, Flemming, and  Gibbs \cite{Flem} constructed a 
hydrodynamic description of solids including  
the new variable $m \equiv \delta\rho/{\bar \rho}+e_1$.  
According to their theory, 
the  {\it vacancy concentration} $c$ 
may  be defined by  
\be 
c= c_{0}- m= 
c_{0}- ( \delta\rho/{\bar \rho}+e_1) 
\label{eq:10}
\en 
where  $c_{0}$ is the average $\av{c}$ 
dependent on $T$ and $\bar \rho$.  On the other hand, 
Granato  claimed relevance  of interstitials 
in amorphous solids because they can greatly 
decrease the shear modulus \cite{Granato}.  
However, because vacancies and interstitials are point 
defects, it may be more appropriate to 
treat $c$  as the 
local free-volume fraction  \cite{Flem,Grest} 
which can take continuous values at each lattice site. 
Then $c_0$ is the average  free-volume fraction 
and we may assume $0 \le c_0 \ll 1$. 
The role of $m$ or $c$ is expected to 
be crucial in amorphous solids.  
We try to include  the  vacancy variable 
in our nonlinear slip model (2D vacancy model),

The dynamic equations obeyed by the mass density 
and the momentum density are of the usual forms, 
\be 
\frac{\p}{\p t} \delta\rho = -{\bar \rho} \nabla\cdot {\bi v}, 
\label{eq:11}
\en 
\be
{\bar \rho} \frac{\p}{\p t}{\bi v} = 
 \nabla\cdot\asigma 
 +\eta_0 \nabla^2{\bi v} + \nabla \cdot \asigma_{\rm R}  .    
\label{eq:12}
\en 
In (12) the force density, the first term   on the right hand side, is 
written in terms of the free energy functional 
$F=F\{\rho, {\bi u}\}$ as 
\be 
\nabla\cdot\asigma= -{\bar \rho}\nabla(\delta F/\delta \rho)_{\bi u}
- (\delta F/\delta{\bi u})_{\rho}= - 
(\delta F/\delta{\bi u})_{m} ,  
\label{eq:13}
\en 
where use has been made of the identities  
 $(\delta/\delta {\bi u})_{\rho}= (\delta/\delta {\bi u})_{m}
- \nabla (\delta/\delta m)_{\bi u}$ 
and $(\delta/\delta m)_{\bi u}= {\bar{\rho}} 
(\delta/\delta \rho)_{\bi u}$. 
The random stress 
tensor $\asigma_{\rm R}$ satisfies (8). 
It is important that, in the presence of  the vacancy field, 
 the lattice velocity 
$\p {\bi u}/\p t$ is different from 
the mass velocity ${\bi v}$ and is assumed to be 
of the form, 
\be 
\frac{\p}{\p t}{\bi u} = {\bi v} - \lambda_0 
\bigg (\frac{\delta}{\delta {\bi u}} F \bigg )_{\rho}  
+ {\bi \zeta}_{\rm R},  
\label{eq:14}
\en 
where $\lambda_0$ is the kinetic coefficient 
and the components  ${\zeta}_\alpha^{\rm R}$ 
of the random force vector  ${\bi \zeta}_{\rm R}$ are 
characterized by 
\be 
\av{\zeta_{\alpha}^{\rm R}({\bi r},t)
\zeta^{\rm R}_{\beta}({\bi r}',t')}=
2k_{\rm B}T \lambda_0 \delta_{\alpha\beta}
\delta({\bi r}-{\bi r}')\delta(t-t').  
\label{eq:15}
\en 
From (12) and (14) the equation for $m$ 
is expressed as 
\be 
\frac{\p}{\p t} m=  \nabla\cdot \lambda_0
\bigg [\nabla \bigg ( 
\frac{\delta }{\delta m} F \bigg )_{\bi u} 
+ \nabla\cdot\asigma \bigg ] + \nabla\cdot{\bi \zeta}_{\rm R}.  
\label{eq:16}
\en 
It is worth noting that 
 (16) is similar to  the dynamic 
equation for the concentration  
in visccoelastic fluid mixtures \cite{Onukibook,Fre}. 
In passing, owing to the special form of (13),  
the time derivative of the 
total free energy $F\{\rho, {\bi u}\} +\int d{\bi r}{\bar{\rho}} v^2/2 $ 
becomes nonnegative-definite in the absence of applied stress  
if the random noises are neglected. 
This is a self-consistent condition 
of  Langevin equations 
ensuring attainment of 
equilibrium \cite{Onukibook}.

The free energy $F=\int d{\bi r}f$ is 
 a functional of $\bi u$ and $\delta\rho$ 
 (or $m$).  We assume the free energy density 
 $f$ in the form,  
\be  
f=  \frac{A}{2} m^2 + \frac{B}{4} m^4 
  +  \alpha m e_1 + 
  +\frac{C}{2}|\nabla \delta\rho|^2+ 
\frac{K_0}{2} e_1^2  + \mu(m) {\cal F}(e_3', e_2'), 
\label{eq:17}
\en 
where $A$, $B$,  and $C$ are positive constants 
and $\alpha$ is a coupling constant. 
The gradient term 
($\propto C$) is introduced to 
 suppress  the density fluctuations with  
  short length scales. 
(We are considering the density fluctuations longer than 
the peak distance of the pair correlation 
function.)    
For simplicity, we neglect  the other  gradient 
terms involving  the gradients of $m$ and the strains.
In our theory it is most important that 
 the shear modulus $\mu(m)$ sensitively depends 
on $m$.  Near glass transitions, 
with decreasing $m$  or increasing $c$, 
$\mu(m)$ is expected to   decrease  abruptly 
from a finite value $\mu_0$  to zero 
($=$ fluid value) around a threshold value of $m$ or $c$.

Analytic calculations of the above  model 
are  difficult except for idealized 
situations. 
As a simple exercise,  
let us consider small deviations 
with wave vector $\bi k$ around a homogeneously 
strained state with $e_3=\gamma$, where the sytem is at rest 
and the harmonic approximation can be made for 
the elastic energy. 
Then the mechanical equilibrium condition 
allows elimination of $\bi u$ from (16) 
\cite{Onukibook}. We  consider the 
 steady-state variance of m in the long wavelength limit, 
$\chi_{\rm v} = \lim_{k\rightarrow 0}\av{|m_{\bi k}|^2}$, in 2D.  
 In the linear approximation it becomes 
 dependent on the angle of $\bi k$: 
\be 
\frac{k_{\rm B}T}{\chi_{\rm v}}= 
A- \frac{\alpha^2}{L_0}   
-\frac{2\alpha }{L_0}\mu_1\gamma \sin 2\varphi - 
\frac{1}{L_0} 
(\mu_1\gamma)^2[1+ (\cos 2\varphi)^2K_0/\mu_0] ,  
\label{eq:18}
\en 
where  $\sin 2\varphi=2k_xk_y/k^2$, 
$\mu(m)= \mu_0+\mu_1m + \cdots$ 
and $L_0=K_0+ \mu_0$.  
The vacancy diffusion constant  
is written as 
$D_{\rm v} = \lambda_0 k_{|rm B}T /\chi_{\rm v}$. 
The steady-state density variance 
$\chi_\rho = \lim_{k\rightarrow 0}\av{|\rho_{\bi k}|^2}/\rho^2$ 
in 2D is written as 
\be 
\chi_\rho= \frac{k_{\rm B}T}{L_0}+ 
\bigg [ 1+ \frac{1}{L_0} 
(\alpha + \mu_1\gamma\sin2\varphi)\bigg ]^2 \chi_{\rm v}.
\label{eq:19}
\en 
The first term is the usual 
term, while the second term arises from the vacancy fluctuations. 
The dependence of 
these quantities on $\gamma$ and the angle $\varphi$ can become 
appreciable with increasing  $\mu_1$ and$/$or decreasing 
$A$.  For $\alpha=0$, for example,  
the homogeneous state is unstable for $\gamma>
\gamma_{\rm c}= (\mu_0 A)^{1/2}/\mu_1$ against  
the flluctuations with ${\bi {k}}$ 
along the $x$ or $y$ axis. 
Note that $\gamma_{\rm c}\ll 1$ 
can hold if $\mu_1/\mu_0 \gg 1$.  
It is easy to confirm that essentially the 
same result follows in 3D in both shear 
and elongational deformations.  
Thus, in elastically deformed 
amorphous solids (before onset of 
plastic flow), we predict enhancement 
of the density fluctuations  
in particular directions 
of the wave vector.

We then show our numerical results of 
our model in (11)-(17) obtained on a $128\times 128$ 
square lattice for the case of applying a constant $\gdot$. 
Again the units 
of  space, time, and strains are 
$\eta_0/(\rho \mu_0)^{1/2}$, 
$\eta_0/\mu_0$,    and 
$\gamma_{\rm p}= 2/\sqrt{3}$, respectively. 
The $m$ is also divided by   $\gamma_{\rm p}$.
The dimensionless mass density  is 
given by 
$\delta\rho/\rho \gamma_{\rm p}$.
Though it is not clear 
what functional form of $\mu(m)$ 
is appropriate, we tentatively use 
\be 
\mu(m) = \mu_0 \exp[- A_0/(m+m_0)] 
\label{eq:20}
\en 
for $m+m_0>0$ and $\mu(m)=0$ for $m+m_0\le 0$.  
Here  $A_0$ is a constant 
and  is set equal to 0.1 in our simulation, 
while $m_0$ is a control 
parameter representing the closeness to the 
glass transition.  In fact, if $\mu(m)$ goes to zero 
at $c= c_{\rm c}$, we have $m_0= c_{c}-c_{0}$  
from (10). However, essentially the same 
numerical results follow as long as 
$\mu(m)$ strongly depends on $m$ 
(for example, from the simpler form $G(m) 
\propto (m+m_0)^2$ instead of (20)). 
 The other parameter values used  are 
$A=C=1$, $B=10$, $\alpha=0$, $\lambda_0=0.1$ 
and $\epsilon=0.1$. While the initial conditions 
for $\bi v$, $\bi u$ and $\theta$ 
are the same as in 
the previous case of Figs.3 and 4, we assign 
random Gaussian randum numbers with variance $0.1$  
to $m$. 
In Fig.5 we show  time evolution of the 
scaled  shear stress as a function of the scaled 
average strain $\gdot t/\gamma_{\rm p}$ at $\gdot=10^{-3}$ 
for $m_0=1,0.6$, and $0.2$.  
Fig.6a and Fig.6b display snapshots of 
$e_3$, $e_2$, $c-c_0=-m$, and $\delta \rho$ 
at $t=250$ and $420$ on the curve of $m_0=0.6$  in Fig.5. 
Fig.7 demonstrates that  these strain-induced 
disordered states are metastable as in Fig.4. 
Salient features are 
as follows:  
(i) Plastic flow is induced earlier 
than in the absence of vacancies, 
because slip formation is easier 
in regions with smaller $m$. Onset of 
plastic flow is   
sensitive to the initial randomness of $m$. 
(ii) The effective viscosity $\eta_{\rm eff}= 
\av{\sigma_{xy}}/\gdot$ becomes 
smaller with decreasing the parameter $m_0$ in (17) 
or increasing the elastic inhomgeneity. 
(iii) The   $\eta_{\rm eff}$ 
also decreases as a function of time on long time scales
 because of slow accumulation of 
vacancies around slips. In our model, though 
this tendency is considerably suppressed by 
the quartic term ($\propto B)$ in (16), 
vacancy accumulation can lead to fluidization     
at long times after percolation of the slips. 
(iii) The density fluctuations  are induced 
and are frozen after cessation of shear.
As demonstared in Fig.7, the fluctuation variances 
of the vacancy and the density 
are much enhanced than in the initial state. 
Owing to the gradient term in (16), 
the density smoothly varies in space 
(compared with $m$).

\section{Concluding remarks}

We have presented a first time-dependent 
Ginzburg-Landau theory 
accounting for 
inhomogeneous nonlinear elastic deformations. 
It is a coarse-grained theory and 
the sharp peak  of the structure factor, 
which is essential in the mode coupling 
theory \cite{mode},  
does not come into play.  
Instead,  periodicity of the elastic energy with respect 
to the strains $e_3'$ and $e_2'$  
and  vacancy-dependence 
of the shear modulus are 
two major ingredients of our theory,  
which bring about  
metastable structural disordered states 
with  heterogeneities in 
the vacancy and mass densities. 
As a closely related effect, 
Fischer \cite{Fischer}
observed excess scattering 
from nearly static density fluctuations 
with sizes in a range of 20-200 nm 
in glass-forming fluids. 
We believe that such large-scale 
frozen fluctuations can arise only as a result of  
nonlinear elastic deformations induced by structural 
disorder.   
We also mention  previous simulations. 
Argon {\it et al}. \cite{Argon} applied a tensile 
(elongational) strain 
to a  2D  model to produce slips and shear bands  making  
angles of $\pm 45^\circ$ with respect to the 
stretched direction in agreement with experiments.  
The same results also follow from our model 
and will be reported shortly. We 
will show that the elastic energy to create a slip under tension 
is minimum for these  directions.   
Ikeda {\it et al.}  \cite{Ikeda}
 applied a tensile strain 
to a 3D model to induce  
a change from a perfect crystal to 
an amorphous solid.

However,  we admit that our results are still  
preliminary and it remains unclear how 
our theory corresponds to  real physical effects.  
In particular, the  initial conditions 
(${\bi u}={\bi 0}$ and $\theta=$const.) 
should be inappropriate for amorphous solids 
(where structural disorder should preexist even before 
applying strain). In future work, we should start with 
 initial states with various  amounts of disorder  and 
should  also examine what 
parameter values are appropriate for 
amorphous solids.

We mention analogous  
and instructive examples. 
In solids 
undergoing phase separation or structural 
 transitions,  
the dependence of the shear modulus on 
the order parameter, written as  $m$,  
in the form $\mu=\mu_0 +\mu_1m$ 
 is of crucial importance 
in phase ordering 
\cite{Onukibook,Onuki-Furukawa}. 
In all  these cases,  local minimization of 
the nonlinear  elastic energy 
($(\delta F/\delta {\bi u})_m= {\bi 0}$) gives rise to 
heterogeneous metastable states  where time-evolution 
is pinned.  In viscoelastic polymer systems,   
the composition-dependence of the viscoelastic properties 
can give rise to shear-induced phase separation 
\cite{Onukibook,Fre}, where the scattered light intensity 
takes a form similar to $\chi_{\rm v}$ in (18).

\vspace{1cm}
{Figure Captions}\\

\no Fig.1, 
Stress vs strain in  units of $\mu_0\gamma_0$ and 
$\gamma_{\rm p}$ 
 obtained from  the discretized 
version of the 1D model (1) 
after application of constant shear rate.  
Here  $\gdot/\omega_0 
= 10^{-3}, 10^{-4}, 10^{-5},$ and $10^{-6}$ from 
above. The inset displays a  snapshot of 
the deviations  
$ u_j/\gamma_0 - \gdot t j$  
($0 \le j \le 600)$  
at $t=720$ for $\gdot=10^{-3}$.\\

\no Fig.2, 
Function  
${\cal F}(e_3,e_2)$ defined by (5), 
which represents  the 2D shear deformation energy density 
divided by $\mu_0$ 
for $\theta=0$. \\

\no Fig.3,  
Stress  vs strain 
 for $\gdot=10^{-3}$ and $10^{-4}$ obtained from 
the 2D slip model (7). In the inset  $e_3$ and $e_2$  at $t=400$ 
for   $\gdot=10^{-3}$ are shown.  \\

\no Fig.4, 
Scaled stress and elastic energy vs time, 
which demonstates freezing of 
strain-induced diordered states 
of  (7) at long times. 
The shear is switched off at 
(a) $t=190$, (b) $t=220$, and (c) $t=440$,  
as indicated by arrows, using the same parameters as for 
the curve of $\gdot=10^{-3}$ in Fig.3.  
 \\

\no Fig.5, 
Stress  vs strain 
 for $\gdot=10^{-3}$  obtained from integration of 
(11), (12), and (14) using 
(16) and (17) at $\gdot=10^{-3}$.  
Here $m_0=1, 0.6$, and $0.2$ from above.\\

\no Fig.6, 
Snapshots of 
$e_3$, $e_2$, $c-c_0=-m$, and $\delta \rho$ 
at  (a) $t=250$  and  (b) $420$  for $\gdot=10^{-3}$ 
on the curve of $m_0=0.6$ 
in Fig.5. \\

\no  Fig.7, 
Freezing of strain-induced diordered states 
of  the  vacancy  model.  Shown are 
the  density variance 
$[\av{(\delta \rho)^2}]^{1/2}/\rho \gamma_p$, 
the vacancy variance 
$[\av{(\delta m)^2}]^{1/2}/\gamma_p$, 
and the average elastic energy 
$\av{f_{\rm el}}$.  As in Fig.4,  
the shear rate  is switched off at 
 $t=220$. The frozen 
 density profile at $t=2\times 10^4$ is 
 shown in the inset.\\

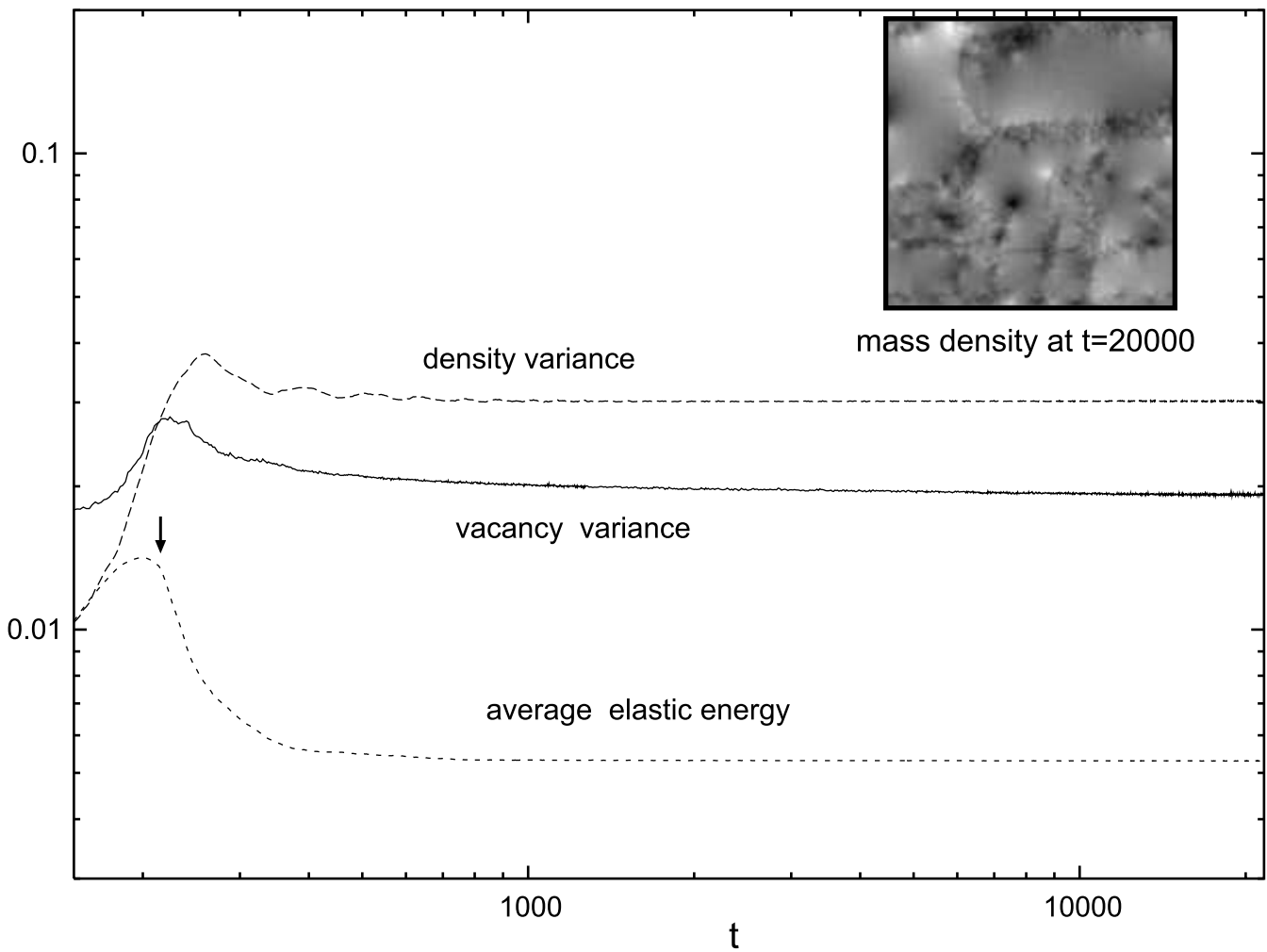
\end{document}